\documentclass{ar-1col-S2O}
\usepackage[numbers]{natbib}
\usepackage{url}
\begin{document}

\markboth{Gopalakrishnan and Vasseur}{Superdiffusion in isotropic magnets}

\title{Superdiffusion from nonabelian symmetries in nearly integrable systems}

\author{Sarang Gopalakrishnan$^1$ and Romain Vasseur$^2$
\affil{$^1$Department of Electrical and Computer Engineering, Princeton University, Princeton NJ 08544, USA; email: sgopalakrishnan@princeton.edu}
\affil{$^2$Department of Physics, University of Massachusetts Amherst, Amherst MA 01003, USA}}

\begin{abstract}
The Heisenberg spin chain is a canonical integrable model. As such, it features stable ballistically propagating quasiparticles, but spin transport is sub-ballistic at any nonzero temperature: an initially localized spin fluctuation spreads in time $t$ to a width $t^{2/3}$. This exponent, as well as the functional form of the dynamical spin correlation function, suggest that spin transport is in the Kardar-Parisi-Zhang (KPZ) universality class. However, the full counting statistics of magnetization is manifestly incompatible with KPZ scaling. A simple two-mode hydrodynamic description, derivable from microscopic principles, captures both the KPZ scaling of the correlation function and the coarse features of the full counting statistics, but remains to be numerically validated. These results generalize to any integrable spin chain invariant under a continuous nonabelian symmetry, and are surprisingly robust against moderately strong integrability-breaking perturbations that respect the nonabelian symmetry. 
\end{abstract}


\maketitle

\tableofcontents

\section{INTRODUCTION}

Charge transport is the canonical experimental probe of many-body dynamics in solid-state systems, and the Hubbard model is a canonical theoretical model for the physics of correlated electrons. Moreover, the one-dimensional Hubbard model is exactly solvable in some sense. It might seem surprising, therefore, that the following very basic question about transport in the Hubbard model---\emph{what is the limiting low-frequency behavior of the a.c. charge conductivity at a fixed nonzero temperature?}---remained even qualitatively unanswered until a few years ago. In part, this regime might have seemed physically irrelevant: at finite temperature and low frequencies, the intrinsic dynamics of the Hubbard model would be modified by coupling to extraneous degrees of freedom like phonons. In part, however, a usable theoretical framework for addressing this question was also lacking. 

The experimental and theoretical advances of the past twenty years have increased both the salience of this regime and our ability to analyze it. On the experimental front, the advent of ultracold atomic gases~\cite{RevModPhys.80.885} made it feasible to study the late-time dynamics of systems that are well isolated despite being at high temperature. Numerical methods based on tensor networks have enabled large-scale simulations of high-temperature dynamics; the phenomena we will discuss was first seen in these simulations, and had not been suspected before they were observed~\cite{1742-5468-2009-02-P02035, vznidarivc2011transport, PhysRevB.80.184402, lzp, bertini2020finite}. On the theoretical side, these advances led to a revival of interest in the dynamics of integrable systems, leading to a cluster of new developments~\cite{PhysRevLett.106.217206, PhysRevLett.115.157201}---most crucially, generalized hydrodynamics (GHD)~\cite{Doyon, Fagotti, doyon2019lecture, Bastianello_2022}---that led to a physical picture of high-temperature transport in integrable systems. 
GHD incorporates the distinctive features of integrable dynamics~\cite{Doyon,  Fagotti, SciPostPhys.2.2.014, PhysRevLett.119.020602,  BBH0, BBH,PhysRevLett.119.020602, piroli2017, GHDII, doyon2017dynamics, solitongases,PhysRevLett.119.195301, PhysRevB.96.081118,PhysRevB.97.081111,alba2017entanglement, PhysRevLett.120.176801, dbd1, ghkv, dbd2, PhysRevB.100.035108,PhysRevLett.122.090601, 2019arXiv190601654B, 10.21468/SciPostPhys.8.2.016, 2019arXiv190807320B, 2020arXiv200407113Y, PhysRevLett.125.070602,10.21468/SciPostPhys.8.3.041}: the presence of infinitely many conservation laws and of stable ballistically propagating quasiparticles. This framework has led to quantitative explanations of many phenomena, including Drude weights~\cite{PhysRevLett.119.020602,BBH,
PhysRevB.96.081118,GHDII}, diffusion constants~\cite{dbd1,ghkv, dbd2,GV19,medenjak2019diffusion,2019arXiv191201551D}
 and the presence of anomalous transport in strongly interacting spin chains~\cite{1742-5468-2009-02-P02035,lzp,sanchez2018anomalous, idmp, GV19,gvw,PhysRevLett.123.186601,PhysRevLett.122.210602, PhysRevE.100.042116, 2019arXiv190905263A,vir2019,2003.05957,PhysRevLett.125.070601,PhysRevB.102.115121}.

At this point, many of the puzzles concerning the hydrodynamics of integrable systems have been resolved. It is understood that transport in these systems is generically ballistic in the long-distance limit, with a Drude weight that has a natural interpretation in terms of charge-carrying quasiparticles (whose effective charge is renormalized by interaction effects). In some cases, however, the renormalization effects are singular, and cancel out the leading ballistic transport \emph{for some charges}. These charges then undergo either normal or anomalous diffusion. Anomalous diffusion occurs, as we will discuss, when both the Hamiltonian and the initial state are invariant under a continuous nonabelian symmetry~\cite{2020arXiv200908425I}, and the transported charge transforms nontrivially under the symmetry. This is the case for charge transport in the Hubbard model at half filling, and for spin transport in the Hubbard and Heisenberg models absent an external magnetic field. 

A surprising feature of this anomalous transport---first observed in simulations of the Heisenberg model~\cite{lzp, PhysRevLett.122.210602}---is that the dynamical spin structure factor follows one of the scaling functions of the Kardar-Parisi-Zhang (KPZ) universality class~\cite{kpz, Prahofer2004}. This correspondence is especially remarkable as the distinguishing features of KPZ physics~\cite{quastel2014kardar} (noise and non-equilibrium conditions) are absent in the linear-response dynamics of integrable spin chains. The anomalous nature of spin transport at the Heisenberg point has been observed in experimental studies of solid-state magnets~\cite{Scheie2021} and ultracold gases~\cite{2107.00038}. At present there is a quantitative theory of the exponent~\cite{GV19, NMKI19, PhysRevLett.125.070601,PhysRevLett.123.186601} (which is argued to be universal for all integrable systems with continuous nonabelian symmetries~\cite{2020arXiv200908425I, 2003.05957, 1909.03799, ye2022universal}), as well as a proposed mechanism for the scaling function~\cite{vir2019}. However, the relation between spin transport in integrable systems and the KPZ equation remains unsettled: however, based on symmetry considerations, it has become clear that there can be no direct correspondence between the spin density and any field in the KPZ equation~\cite{PhysRevLett.128.090604, de2022non}. It was recently proposed that the spin dynamics maps onto a modified two-mode KPZ equation~\cite{de2022non}; this proposal circumvents the symmetry obstructions, but remains to be verified by numerics or experiment.

In the rest of this article we survey what is known about anomalous transport in integrable systems and its relation to the KPZ equation, and what remains mysterious. For specificity (and to make contact with numerics and experiment), we will focus on spin transport in the Heisenberg model. However, all our theoretical claims extend to other integrable systems with nonabelian symmetries. We will focus on recent theoretical advances at the isotropic Heisenberg point. Many of the technical points that we have omitted are treated in the more detailed reviews~\cite{bertini2020finite, Bulchandani_2021, gopalakrishnan2022anomalous}.

\section{CONTEXT}

We begin with a very brief overview of some of the background results and ideas that provide the context for recent work on anomalous transport. First, we introduce ideas from generalized hydrodynamics that will be relevant for our discussion of the Heisenberg spin chain. Second, we introduce the transport phase diagram of the anisotropic XXZ spin chain, identifying the Heisenberg model as a sort of dynamical critical point. Finally, we review the basic dynamical features of the KPZ universality class in one spatial dimension; these will form the basis for subsequent comparisons with the behavior of the Heisenberg spin chain.

\subsection{Generalized hydrodynamics (GHD)}

A defining feature of integrable systems is that they possess stable quasiparticles~\cite{Takahashi}. The eigenstates of integrable systems can be labeled by their quasiparticle content. In terms of these quasiparticle labels, one can straightforwardly compute thermodynamic properties of integrable systems, as these depend purely on the \emph{eigenvalues} of the Hamiltonian, which can be found by numerically solving the coupled Bethe equations. However, quantities such as the conductivity, which is given by the Kubo formula (see {\it e.g.} Ref.~\cite{bertini2020finite})
\begin{equation}\label{kubo}
 \sigma(\omega) = \beta \int_0^\infty dt \int dx {\rm e}^{i \omega t} \langle j(x,t) j(0,0)  \rangle,
\end{equation}
written here in the high temperature limit $\beta \to 0$ for simplicity,
depend on matrix elements of the local current operators $j$ between the \emph{eigenvectors} of an integrable Hamiltonian. Formal expressions exist in some cases for these quantities~\cite{dugave2016thermal}, but are generally unwieldy, and their large-system asymptotics has proven difficult to extract. 

The GHD framework approaches integrable dynamics from a very different perspective. Let us begin with an integrable lattice model in the thermodynamic limit. This system contains an infinite number of conserved charges, of the form $Q_n = \sum_i q^n_i$, where $q^n_i$ is a \emph{local} operator, and the label $i$ runs over all sites in the chain. Because of these conserved charges, the space of equilibrium states of an integrable system is much larger than that of a generic system: the equilibrium states are called generalized Gibbs ensembles (GGEs), and can be specified in terms of density matrices of the form $\rho \propto \exp(-\sum\nolimits_n \mu_n Q_n)$, where we have fixed a chemical potential for each conserved charge. Equivalently, one can specify equilibrium states in terms of the distribution function of stable quasiparticles. These two descriptions are related by a transformation called ``string-charge duality''~\cite{1742-5468-2016-6-063101}. 

In the spirit of hydrodynamics, we divide the system into cells of mesoscopic size $\ell$; each cell is labeled by its position $x$, which we will treat as a continuous index. In the initial state, each cell is in an equilibrium state, but the chemical potentials are slowly spatially varying, on length-scales much longer than $\ell$, so the state can be written in the form 
\begin{equation}\label{hydroform}
\rho \propto \exp\left(-\sum_n \int dx \mu_n(x) q_n(x)\right).
\end{equation}
The key assumption of GHD is that local relaxation to a GGE is much faster than large-scale charge rearrangements. Therefore the time-evolving state can continue to be written in the form~(\ref{hydroform}), and to compute charge dynamics it suffices to compute the time evolution of the chemical potentials or (equivalently) the local quasiparticle distribution. This quasiparticle distribution evolves in time as quasiparticles propagate ballistically through the system (with a state-dependent effective velocity). At any time, the instantaneous charge profile can be inferred from the local quasiparticle distribution using string-charge duality. Crucially, this procedure allows one to compute transport in integrable systems with no direct reference to matrix elements.

\subsection{Anisotropic XXZ model}

Naively, one might expect that transport in integrable systems should always be ballistic because of the presence of ballistic quasiparticles. This expectation fails in one of the simplest integrable lattice models, the anisotropic XXZ spin chain:
\begin{equation}\label{xxzmodel}
H_{\mathrm{XXZ}}(\Delta) = \sum\nolimits_i (S^x_i S^x_{i+1} + S^y_i S^y_{i+1} + \Delta S^z_i S^z_{i+1}).
\end{equation}
The XXZ model is a generalization of the Heisenberg model (which occurs at $\Delta = 1$). Remarkably, this model undergoes a dynamical phase transition precisely at the Heisenberg point, at which spin transport changes from ballistic ($|\Delta| < 1$) to diffusive ($|\Delta| > 1$). In the sense we are using here, a dynamical phase transition occurs when the late-time limit of a dynamical correlation function changes singularly as some parameter is tuned. Specifically, in the XXZ model~(\ref{xxzmodel}), the late-time limit of the spin-current correlation function, 
\begin{equation}\label{drudeW}
\lim\nolimits_{t \to \infty} \lim\nolimits_{L \to \infty} \langle J(t) J(0)\rangle,
\end{equation}
with $J = \int dx j$ the total spin current, is identically zero when $|\Delta| \geq 1$ and nonzero when $|\Delta| < 1$~\cite{PhysRevLett.106.217206,PhysRevLett.111.057203}. In systems with ballistic transport, the limit~(\ref{drudeW}) is nonzero because initially injected currents persist forever. Therefore the transition at $\Delta = 1$ is associated with the vanishing of ballistic spin transport. However, even for $|\Delta|>1$, the Hamiltonian~(\ref{xxzmodel}) hosts stable ballistic quasiparticles~\cite{bertini2020finite}: energy transport remains ballistic, and so does spin transport starting from spin-polarized initial states. This coexistence of ballistic quasiparticles with sub-ballistic transport is an interesting consequence of the interplay between integrability and symmetries, as we will discuss below.

It might seem strange that a quantity like Eq.~(\ref{drudeW}) can change in a singular way as one continuously tunes a parameter in the Hamiltonian. However, such singular changes also happen in generic weakly interacting Fermi gases in dimensions greater than one. The strictly noninteracting gas is integrable, and has persistent currents, so the limit~(\ref{drudeW}) is nonzero. However, any nonzero interaction strength or disorder leads to the relaxation of currents, so the limit~(\ref{drudeW}) is zero, although the \emph{time} the system takes to approach the limiting value diverges as one approaches the noninteracting point. As this example illustrates, singular limits of the form~(\ref{drudeW}) are not unexpected. What is more unexpected is that this behavior happens for particular charges (in this case, the spin) as one tunes through a family of integrable models. 

\subsubsection{Easy-axis regime: absence of ballistic transport}

The vanishing of ballistic transport is most clearly illustrated in the $\Delta = \infty$ limit at infinite temperature. (We take the infinite-temperature limit first.) In this limit, one obtains a constrained Hamiltonian, the ``folded XXZ model''~\cite{zadnik2021folded, fagotti2014conservation, pozsgay2016real}:
\begin{equation}\label{foldedxxz}
H_{\mathrm{folded}} = J \sum\nolimits_i (1 + 4 S^z_{i-1} S^z_{i+2}) (S^x_i S^x_{i+1} + S^y_i S^y_{i+1}).
\end{equation}
In the folded XXZ model, spins can only move if their motion does not change the total number of domain walls. The quasiparticles of this model are magnons, which move ballistically, and a frozen background of spins, which have no intrinsic dynamics. As magnons move, they scatter off the frozen background. These scattering processes lead to the screening of the magnon spin (since it continually flips as it propagates through the random-spin frozen background), and also to the Brownian motion of the frozen pattern, which leads to diffusive spin transport despite ballistic quasiparticle motion. Note that spin transport is only diffusive when the frozen background is precisely unmagnetized; otherwise, magnons do carry some net magnetization and give ballistic transport. This picture leads to a number of nontrivial consequences that have been numerically confirmed, such as the anomalous decay of local autocorrelation functions away from half filling~\cite{gvw}, the anomalous response to weak integrability-breaking perturbations~\cite{2021arXiv210913251D}, the strongly non-gaussian full counting statistics of magnetization~\cite{2022arXiv220309526G}, and the arrested growth of spin fluctuations~\cite{squensembles}.

Away from $\Delta = \infty$, this picture continues to apply in many respects. However, the frozen pattern is now no longer perfectly frozen. Instead, it consists of spin domains of various sizes. A domain of size $n$ moves at $n$th order in perturbation theory, and its velocity is correspondingly exponentially suppressed. These large domains do not affect the low-frequency conductivity, though they do affect the local autocorrelation function~\cite{gvw}. An exact formula for the diffusion constant, including domains of all sizes, can be derived~\cite{GV19, dbd2}. (See also previous related work on the low-temperature behavior~\cite{PhysRevLett.95.187201, PhysRevB.57.8307}.) As $\Delta \to 1^+$, this formula predicts that the spin diffusion constant diverges as $D(\Delta) \sim (\Delta - 1)^{-1/2}$.

When $|\Delta| < 1$, the structure of the quasiparticle spectrum is completely different, as the bound states we have been discussing are no longer stable in general. The quasiparticle hierarchy for $|\Delta| < 1$ is complex and fractal in ways that are peripheral to our discussion; see Refs.~\cite{PhysRevLett.106.217206, PhysRevB.97.081111, PhysRevLett.122.150605, 2019arXiv190905263A} for details. The fact that spin transport in this regime remains ballistic is nontrivial; perhaps the simplest way to understand why this result holds is that in this regime there are quasilocal conserved charges that have a nontrivial overlap with $S^z$~\cite{PhysRevLett.106.217206}. This observation implies that at least part of the spin current is conserved, guaranteeing ballistic spin transport.

\subsection{KPZ and Burgers equations}

The Kardar-Parisi-Zhang (KPZ) equation in one dimension has the form~\cite{kpz}:
\begin{equation}
\partial_t h(x,t) = c + \nu \partial_x^2 h(x,t) + (\lambda/2) (\partial_x h(x,t))^2 + \eta(x,t),
\end{equation}
where $h(x,t)$ is called the ``height field,'' $\eta(x,t)$ is spatiotemporally uncorrelated white noise, and the coefficients $c, \nu, \lambda$ are, respectively, the mean growth rate, the diffusion constant, and the strength of the nonlinearity. The KPZ equation was first explored in the context surface growth, where $h$ is the height of a growing interface subject to random deposition and the diffusion of deposited particles. However, the KPZ equation is ubiquitous in nonequilibrium problems: it describes the generic coarse-grained dynamics of scalar fields in noisy systems with no symmetry constraints beyond spatial inversion, see for example~\cite{corwin2011kardarparisizhang,Spohn2014,Nahum2017,Fontaine2022,PhysRevLett.125.040603,2204.00070,Bernard2020,PhysRevLett.128.070401,PhysRevLett.124.236802}. 

In one dimension, the KPZ equation can be related to the Burgers equation,
\begin{equation}\label{burgers}
\partial_t u(x,t) + u(x,t) \partial_x u(x,t) = \nu \partial_x^2 u(x,t) - \lambda \partial_x \eta(x,t), 
\end{equation}
via the change of variables $u(x,t) = -\lambda \partial_x h(x,t)$. We will refer to $u(x,t)$ as the Burgers field. Unlike the height field, the Burgers field is a conserved density, i.e., $\int dx u(x,t)$ is conserved under Eq.~\ref{burgers}. The Burgers equation describes the nonlinear dynamics of an incompressible fluid in one dimension. If we consider small fluctuations about a state of constant $u$, so that $u(x,t) = u_0 + \tilde u(x,t)$, Eq.~\ref{burgers} describes ballistic motion with a density-dependent velocity. As such, it arises very naturally for interacting models with ballistic motion. The lhs of the Burgers equation generates shocks, which are regularized by the diffusion term on the rhs. The diffusion term in turn entails the noise term in order to maintain an equilibrium state with thermal fluctuations. It is known that the infinite-temperature state is a stationary distribution under Eq.~\ref{burgers}. 

A few observations follow directly from the definitions above. First, under the Burgers equation, density excesses move right while density deficits move left. Since dynamical correlation functions involve \emph{squares} of the density fluctuation, these are left-right symmetric. On the other hand, if one measures the distribution of density transferred across a particular point (i.e., the change in the KPZ height field at that point, $h(x,t) - h(x,0)$), this distribution will generally be skewed~\cite{PhysRevLett.104.230602, amir2011probability, PhysRevLett.106.250603, imamura2013stationary}. Second, fluctuations in the Burgers equation spread with dynamical scaling exponent $z = 3/2$ (i.e., $x \sim t^{2/3}$). A heuristic argument for this is as follows: in equilibrium, the typical value of $u$, coarse-grained over a region of size $\ell$, scales as $1/\sqrt{\ell}$. This is also the characteristic velocity at which fluctuations traverse this region. Therefore, the time taken for a fluctuation to traverse a region of size $\ell$ is $t(\ell) \sim \ell/(1/\sqrt{\ell}) \sim \ell^{3/2}$. 

In addition to the scaling exponents, the scaling functions of the KPZ equation are exactly known for many quantities. For example, spatio-temporal correlations of the Burgers field above an equilibrium state follow the scaling form $C(x,t) = t^{-2/3} f_{\mathrm{KPZ}}(x/t^{2/3})$, where $f_{\mathrm{KPZ}}$ is clearly non-gaussian and has been tabulated in Ref.~\cite{Prahofer2004}. Unlike other scaling functions associated with anomalous transport (e.g., L\'evy flights), $f_{\mathrm{KPZ}}$ falls off parametrically \emph{more} rapidly than a gaussian in the far tails. In addition, the probability distribution $P(h(x,t) - h(x,0))$ (where $h$ is the KPZ height field) is known exactly for various initial conditions, including the equilibrium initial condition~\cite{PhysRevLett.104.230602, amir2011probability, PhysRevLett.106.250603, imamura2013stationary}. Using the relation between the KPZ and Burgers equations this quantity can be understood as the full probability distribution of the charge crossing a particular cut as a function of time. These universal limiting distributions share the feature that they are skewed: in this respect they are very different from the typical distribution functions of conserved charges in equilibrium systems~\cite{RevModPhys.87.593, mcculloch2023full}. 

\section{LINEAR RESPONSE OF HEISENBERG CHAINS}

Most studies of transport in the Heisenberg spin chain have focused on the linear-response a.c. conductivity, governed by the Kubo formula~(\ref{kubo}). Because the thermodynamics of one-dimensional spin chains is qualitatively the same at any finite temperature, it is convenient to take $T \to \infty$ and compute the infinite-temperature limit of the rescaled conductivity $T\sigma(\omega)$. 
The $T \to \infty$ limit can be taken at constant net magnetization, leading to the density matrix $\rho(\mu) \propto \exp(-\mu \sum\nolimits_i S^z_i)$. The standard infinite-temperature state occurs at $\mu = 0$, but it will sometimes be helpful to work at $\mu \neq 0$ and take the limit at the end.

\begin{figure}
 \label{fig:solitons} 
\begin{centering}	\includegraphics[width=0.55\textwidth]{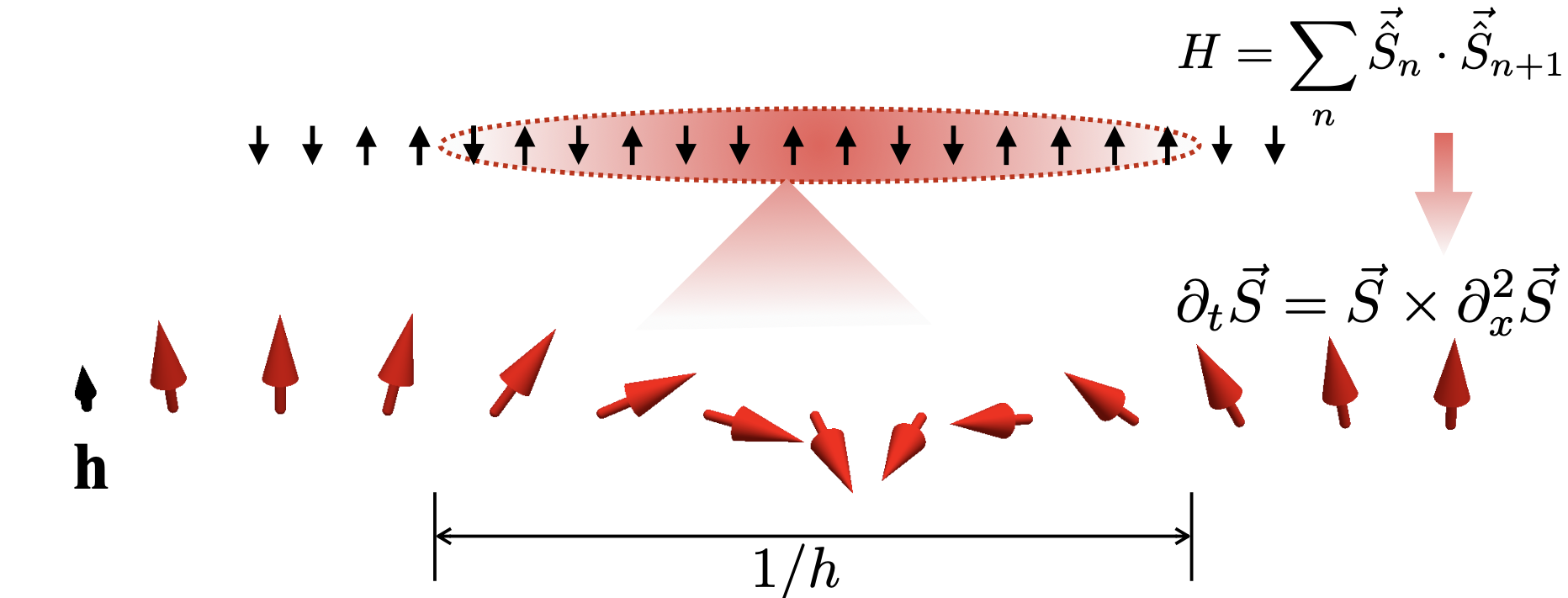}
\end{centering}
	\caption{ {\bf Goldstone solitons -- }  The ``giant'' quasiparticles responsible for superdiffusive transport have a semiclassical description as ``Goldstone solitons'': they correspond to configurations in which the vacuum orientation rotates by an $O(1)$ angle over a length-scale $s$ and then rotates back. These excitations are classical in nature and are soft (low-energy) solutions to the Landau-Lifshitz equations. In the presence of a small magnetic field $h$, transport is dominated by quasiparticles of size $\sim s$.
 Figure reproduced from Ref.~\cite{PhysRevLett.125.070601}.
            }             
\end{figure}

\subsection{Quasiparticle structure}

Following the GHD prescription, we will express the conductivity of the Heisenberg chain in terms of stable quasiparticles.
The Heisenberg chain has infinitely many species of quasiparticles, indexed by an integer $s$. For both ferromagnetic and antiferromagnetic chains, quasiparticles are defined above a reference ferromagnetic vacuum (e.g., all spins pointing down). At $\mu = 0$ the choice of vacuum is arbitrary, giving our description a ``gauge'' redundancy~\cite{vir2019}, but $\mu \neq 0$ fixes a vacuum orientation. Above this vacuum, the elementary quasiparticles are magnons (i.e., single spin-flips) and quasiparticles of species $s > 1$ (also called $s$-strings) are bound states of $s$ magnons. In the large-$s$ limit, these quasiparticles have a semiclassical description as ``Goldstone solitons''~\cite{PhysRevLett.125.070601}---i.e., configurations in which the vacuum orientation rotates by an $O(1)$ angle over a length-scale $s$ and then rotates back~\cite{vir2019, lrt, Theodorakopoulos1991}. This semiclassical description can be made precise through a mapping to the lattice Landau-Lifshitz model~\cite{PhysRevLett.125.070601}, see Fig.~\ref{fig:solitons}. One can regard Goldstone solitons as wavepackets of size $\sim s$ made up of Goldstone modes of the ferromagnet; as usual with solitons, the effects of nonlinearity and dispersion balance each other, giving rise to a stable excitation. 

The properties of $s$-strings can be derived at arbitrary $s$ using Bethe ansatz techniques. However, in the large-$s$ limit these properties have a natural interpretation in terms of Goldstone modes. Recall that the Goldstone modes of a Heisenberg ferromagnet disperse quadratically. So an $s$-string carries magnetization $\sim s$ and energy $\sim 1/s$ (i.e., energy density $\sim 1/s^2$ over size $s$), and it moves with a characteristic velocity $\sim 1/s$. Moreover, when a soliton encounters a much bigger soliton, it encounters it as a local change of the vacuum orientation; as such, the spin orientation of the smaller soliton changes when it is passing through a larger soliton. 

At small finite net magnetization $\propto \mu$, the density of very large strings goes as $\rho_s \sim \exp(-\mu s)$. The density of strings of size $s \ll 1/\mu$ is essentially $\mu$-independent and scales as $\rho_s \sim 1/s^3$. This behavior was found using Bethe ansatz techniques~\cite{idmp} but can be deduced from the following scaling argument. First, note that a region of size $\ell$ has thermal fluctuations in its magnetization density of order $1/\sqrt{\ell}$. For this region to resolve a net magnetization $\mu$, we require $\mu > 1/\sqrt{\ell}$, i.e., $\ell > 1/\mu^2$. To reconcile this crossover with the crossover between magnetized and unmagnetized strings, we require that a region of size $\ell(\mu) \sim 1/\mu^2$ must have $O(1)$ strings of size $s > 1/\mu$. Therefore, $\sum\nolimits_{s' \geq s} \rho_{s'} \sim 1/s^2$, giving the quoted asymptotics for $\rho_s$.

\subsection{Dynamics of screening and transport}

We now expand the autocorrelation function of the current in terms of contributions due to quasiparticles:
\begin{equation}\label{stringexp}
\langle J(t) J(0) \rangle/L = \sum\nolimits_s \int d\lambda \rho_{s,\lambda} \langle j_{s,\lambda}(t) j_{s,\lambda}(0) \rangle,
\end{equation}
with $L$ the system size. 
Here, $\lambda$ indexes the quasimomenta (rapidities) of each species. The time-dependent current can be decomposed into a conserved and a decaying part $j_{s,\lambda}(t) = \overline{j_{s,\lambda}} + \delta j_{s,\lambda}(t)$. At $\mu \neq 0$, strings never completely depolarize, so Eq.~(\ref{stringexp}) has a nonzero late-time limit, the spin Drude weight $\mathcal{D}$: 
\begin{equation}\label{hdrude}
\mathcal{D} = \lim_{t \to \infty} \langle J(t) J(0) \rangle/L = \sum_s \int d\lambda \, \rho_{s,\lambda} \left\langle \overline{j_{s,\lambda}}^2 \right\rangle . 
\end{equation}
%
%
This quantity is straightforward to compute in GHD; the result~\cite{idmp} is that $\mathcal{D}(\mu) \sim \mu^2 \log \mu$ for small $\mu$. Therefore, as $\mu \to 0$, the spectral weight of the current autocorrelator is almost entirely in the decaying part, which we will now discuss. To simplify the discussion, we will omit the label $\lambda$ (as the integrals over $\lambda$ just add $O(1)$ factors that are not crucial to our discussion). 

\begin{figure}
 \label{fig:diffusionconstant} 
\begin{centering}	\includegraphics[width=0.5\textwidth]{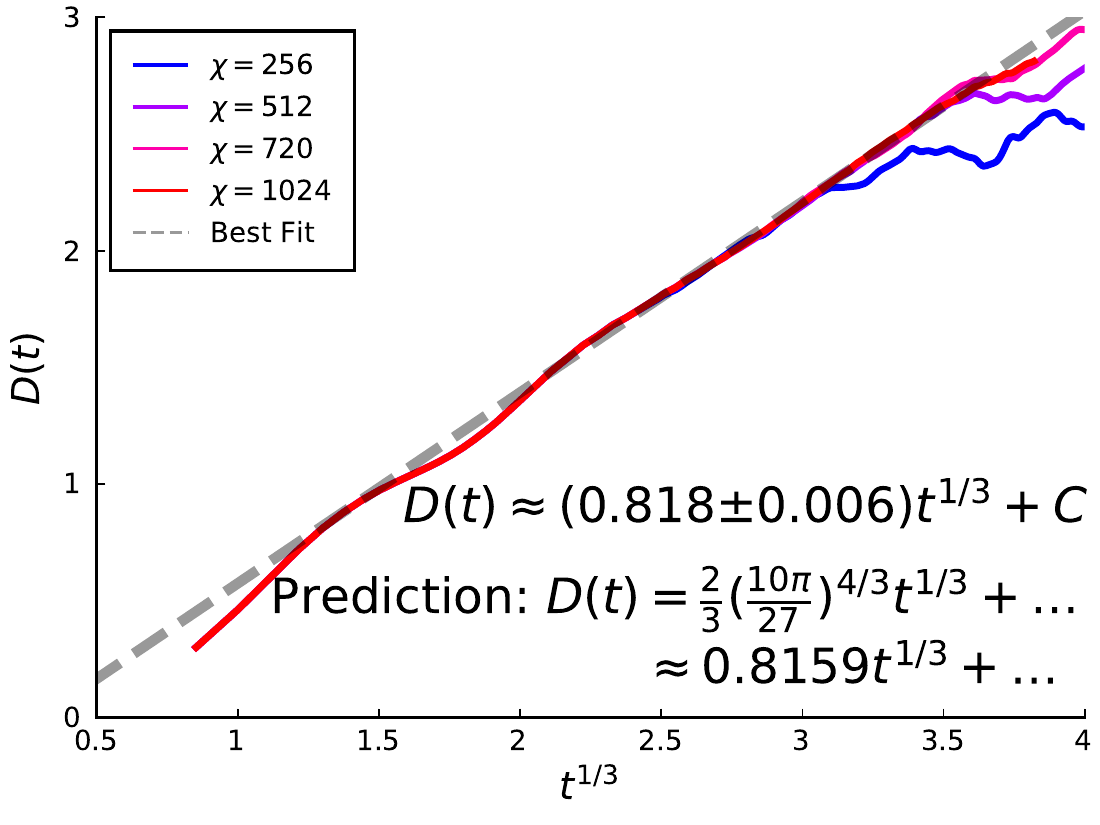}
\end{centering}
	\caption{ {\bf Time dependent diffusion constant} -- Anomalous growth of the time-dependent diffusion constant $D(t) = \frac12 \frac{d }{d t}\sigma^2(t)$, where $\sigma^2(t)$ is the spatial variance of the spin structure factor.  The variance of the spin profiles has been predicted analytically~\cite{PhysRevLett.125.070601} to scale as $\sigma^2(t) \sim \frac12 \left(\frac{10 \pi}{27} \right)^{4/3} t^{4/3}$, in good agreement with tensor-network numerics ($\chi$ is the bond dimension of the matrix-product operators used in the numerics). 
 Figure reproduced from Ref.~\cite{PhysRevLett.127.057201}.
            }             
\end{figure}

At short times, no screening has yet taken place, so $\langle \delta j_s(t) j_s(0)\rangle \sim \langle j_s(0)^2 \rangle \sim (m_s v_s)^2 (1-n_s)$. Here, $m_s = s$ is the bare magnetization of the $s$-string, $v_s \sim 1/s$ is its bare velocity, and $n_s$ is its Fermi occupation factor~\cite{doyon2019lecture}. At very long times, string $s$ is screened by its interaction with larger strings, so $j_s(t) \sim j_s(0) \exp(-t/\tau_s)$, where $\tau_s$ is the characteristic time it takes $s$-strings to decay\footnote{We are assuming exponential decay for simplicity but the argument also works for other functions $f(t/\tau_s)$ that vanish sufficiently fast at large argument.}. One can estimate $\tau_s$ by noting that a string decays whenever it encounters a larger string~\cite{PhysRevLett.127.057201}. As we noted above, the density of strings larger than $s$ scales as $1/s^2$, and the velocity of the $s$-string scales as $1/s$, so $\tau_s \sim s^3$. Putting these expressions together we find that
\begin{equation}
\langle J(t) J(0) \rangle_{\mu = 0}/L \sim \sum\nolimits_s s^{-3} \exp(-t/s^3) \sim t^{-2/3}.
\end{equation}
This result implies that the a.c. conductivity scales as $\sigma(\omega) \sim \omega^{-1/3}$, which in turn implies the dynamical scaling relation $x \sim t^{2/3}$ as was first numerically observed in Refs.~\cite{vznidarivc2011transport}.

Although we introduced these results using heuristic scaling arguments, we emphasize that by exploiting our detailed knowledge of the quasiparticle structure from GHD, one can make these arguments fully quantitative, obtaining prefactors that are in good agreement with numerics~\cite{PhysRevLett.125.070601,PhysRevLett.127.057201} (See Fig.~\ref{fig:diffusionconstant}).

\subsection{Crossovers at finite field and low temperature}

When $\mu \neq 0$, no string gets completely depolarized. However, when $\mu$ is sufficiently small, the conserved part of the current due to strings with $s \ll 1/\mu$ is much smaller than the non-conserved part. Moreover, the weak field does not change the density of such strings. The contribution of each string with $s \ll 1/\mu$ is therefore a regular function (e.g., a Lorentzian) of width $1/\tau_s^3$ and total spectral weight $O(1)$, plus a delta-function contribution with weight of order $\mu^2$. On the other hand, strings with $s \gg 1/\mu$ are not screened, as the population of yet larger strings is strongly suppressed. Therefore, these large strings contribute primarily to the Drude peak, but with the weight of each contribution suppressed as $\exp(-\mu s)$. 

This gives the following crossover behavior for the a.c. conductivity at finite fields. For a fixed $\mu$ the crossover between small and large strings happens at frequency $\omega(\mu) \sim \mu^3$. Above this crossover scale, the a.c. conductivity is largely unaffected by the field. The singular spectral weight below this scale migrates to the Drude peak at $\omega = 0$. Thus the d.c. limit of the conductivity is set by $\omega^{-1/3}|_{\omega \to \mu^3} \sim 1/\mu$. This result can also be computed directly using GHD methods~\cite{NMKI19, gvw}. 

The quasiparticle picture outlined above also naturally extends to low temperatures. This case was studied in detail in Ref.~\cite{PhysRevLett.127.107201}. Since the energy of an $s$-string scales as $1/s$, low temperatures suppress the population of \emph{small} strings (in contrast to fields, which suppress \emph{large} strings). At a low temperature $T$, strings of size $s \ll 1/T$ are absent, while strings much larger than that scale persist with density $\rho_s \sim 1/s^3$. The associated crossover length-scale is simply the thermal correlation length, $\xi \sim 1/T$: below this scale the system looks magnetically ordered and there is no screening. Since the asymptotic long-time behavior is set by the largest strings, transport at all nonzero temperatures remains asymptotically superdiffusive. However, since a magnon must travel at least one thermal correlation length to be screened, this superdiffusion matches on to a short-time ballistic behavior. At a fixed low $T$, transport is ballistic to a distance scale $1/T$ and superdiffusive past that scale. Equivalently, anomalous transport only occurs at frequencies $\omega \leq T$. Thus the prefactor of the low-frequency conductivity diverges at low temperatures: enforcing continuity at the ballistic-superdiffusive crossover, we find that $\sigma(\omega) \sim (\omega T^2)^{-1/3}$ at low frequencies.

\begin{figure}
 \label{fig:KPZ} 
\begin{centering}	\includegraphics[width=0.8\textwidth]{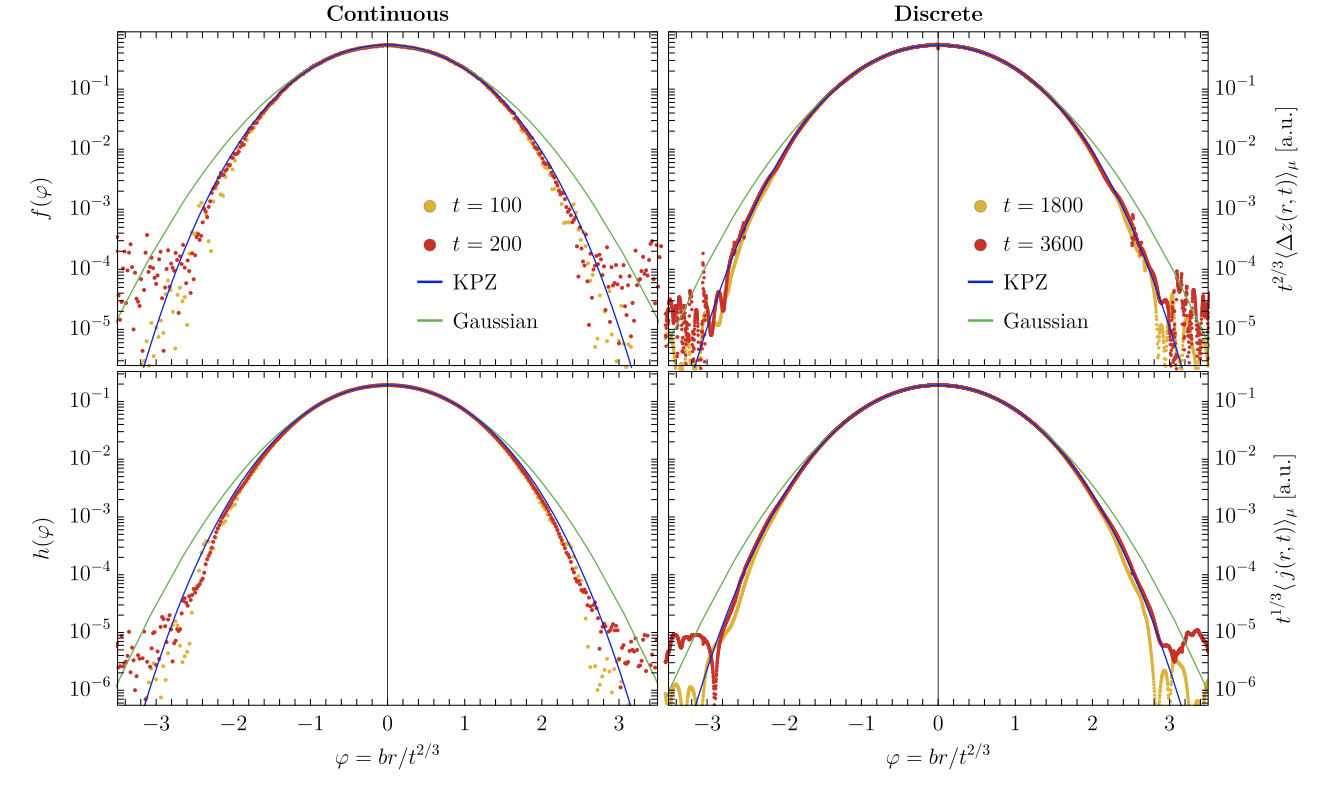}
\end{centering}
	\caption{ {\bf KPZ scaling of correlators} -- Rescaled dynamical correlators in the integrable quantum Heisenberg magnet, with continuous (left) or discrete (right) time. The tails are non-Gaussian, and compatible with a KPZ scaling function. 
 Figure reproduced from Ref.~\cite{PhysRevLett.122.210602}.
            }             
\end{figure}

\subsection{Dynamical structure factor}

So far, we have focused on the a.c. conductivity, i.e., the response to a uniform but time-dependent driving field. One of the most surprising numerical results concerning transport~\cite{PhysRevLett.122.210602}, however, was the discovery that the infinite-temperature dynamical correlation function $\langle S^z(x,t) S^z(0,0) \rangle$ precisely matches the dynamical correlation function of the Burgers field~\cite{PhysRevLett.122.210602}: i.e., $C_{zz}(x,t) = \langle S^z(x,t) S^z(0,0) \rangle = t^{-2/3} f_{\mathrm{KPZ}}(x/t^{2/3})$, where $f_{\mathrm{KPZ}}$ is a function that is known exactly (Fig.~\ref{fig:KPZ})~\cite{Prahofer2004}. The function $f_{\mathrm{KPZ}}$ can be distinguished from a gaussian by the fact that it falls off more rapidly in the far tails, as $f_{\mathrm{KPZ}}(x) \sim e^{-x^3}$. The conductivity is related to the second spatial moment $\int dx\,x^2 C_zz(x,t)$. At present, no ``direct'' calculation of this correlation function (or indeed of any of its higher spatial moments) has been performed within the GHD framework.

\subsection{Generalization beyond Heisenberg}

So far we have focused on the Heisenberg model. However, the only nontrivial dynamical facts we used were (i)~the existence of quadratically dispersing Goldstone modes, which holds in any ferromagnet with a continuous nonabelian symmetry, and (ii)~the stability of quasiparticles, which holds in any integrable system. Therefore, we might expect these results to extend to other integrable systems with a continuous nonabelian symmetry. Extensive numerical evidence~\cite{dupont_moore, 2020arXiv200908425I, PhysRevB.102.115121, 2003.05957, 1909.03799, PhysRevLett.129.230602} supports these claims for classical and quantum spin chains, in both discrete and continuous time, invariant under many different nonabelian symmetries. In all cases where this has been checked, the dynamical structure factor appears to match the KPZ function~\cite{PhysRevLett.129.230602}.


\section{TESTING KPZ BEYOND LINEAR RESPONSE}

Given the analysis of two-point dynamical correlation functions in the Heisenberg spin chain, it is tempting to identify $S^z_i$ (or some coarse-grained version of it) with the field in the Burgers equation. Heuristically, this identification might seem plausible because in the presence of a background magnetization $h$, transport is dominated by quasiparticles of size $h$, which move with speed $\sim h$, giving the Burgers nonlinearity $h \partial_x h$~\cite{NMKI19, gvw}. However, this identification is untenable because the two variables have incompatible symmetries. Even ignoring the fact that spin is a vector and the Burgers field is not, one has to negotiate the much more substantial issue that the Burgers equation is intrinsically \emph{chiral} while the Heisenberg spin chain is not. 

\subsection{Absence of chirality}

Under the Burgers equation, local fluctuations of positive (negative) density move right (left). This chirality is a natural feature of most settings in which the Burgers equation arises. In canonical examples of the Burgers/KPZ universality class, such as the asymmetric exclusion process, $t^{2/3}$ scaling occurs in the dynamics about a ballistically moving front. These problems inherently break inversion and time-reversal symmetry. The Heisenberg model in equilibrium does not. This distinction does not affect the two-point function, since the two-point function is insensitive to the sign of fluctuations. A clear and measurable observable in which this distinction does matter, however, is the ``full counting statistics'' (FCS), i.e., the probability distribution of the number of particles moving across a particular site as a function of time.

We now discuss the behavior of the FCS using the conceptually simplest setup. Consider an initial state on a chain of length $L$ that has a fixed number of particles in each half-system of $L/2$ sites, but is otherwise random. We evolve this state under the dynamical rules of interest and measure the total number of particles that are left of the central bond. We call this $Q(t)$. We focus on times $t \ll L$, so that boundary effects can be ignored. If the dynamics were governed by a Burgers equation, it is easy to see that $Q(t)$ is precisely the change in the KPZ height field at the central bond, $Q(t) \leftrightarrow \delta h(t) \equiv h(0,t) - h(0,0)$. (Recall that the Burgers field $v$ and the KPZ height field $h$ are related by $v(x) = \partial_x h(x)$.) For the KPZ universality, the full probability distribution $P(\delta h(t))$ is exactly known; for equilibrium initial conditions of the Burgers field, the limiting form of $P(\delta h(t))$ is the Baik-Rains distribution~\cite{BR}. Crucially, the Baik-Rains distribution has an $O(1)$ skewness. However, in the Heisenberg model, both the initial state and the evolution we presented were precisely inversion-symmetric, so $P(Q(t))$ must have zero skewness in equilibrium and the magnetization \emph{cannot} simply map on to the Burgers field.

\subsection{Modified KPZ ansatz}

An obvious way to build a nonchiral model out of a chiral one is to make two copies of it with opposite signs of the chirality. When the copies are decoupled, the two-point correlation functions will manifestly be given by $f_{\mathrm{KPZ}}$ but the distribution of $Q(t)$ will be symmetric. Generically, however, the two copies will couple, giving rise to a nontrivial modification of KPZ behavior. We now discuss a minimal set of two coupled equations, in terms of the spin density $\vec{S}$ and a second field $\phi$, that resolves the obvious symmetry issues identified above. 

The first equation is a continuity equation for $\vec{S}$:
\begin{equation}\label{2mode1}
\partial_t \vec{S} + \partial_x (\phi \vec{S}) = D_s \partial_x^2 \vec{S} + \partial_x \vec{\xi} + \ldots
\end{equation}
The rhs contains diffusion and (density-conserving) noise, which are subleading to the effects we will consider here. The content of Eq.~(\ref{2mode1}) is that the scalar field $\phi$ acts as a velocity for spin fluctuations. By symmetry we require $\phi$ to be odd under spatial inversion and under time-reversal. Physically, one can interpret $\phi$ as being related to the imbalance between left and right moving giant quasiparticles. We note that $\phi$ has the same symmetries as the energy current: this is physically sensible as an energy current causes spin fluctuations to drift in the opposite direction through collisions~\cite{Jacopoetal,2022arXiv221203696D}. 

We now write down an equation for $\phi$. In principle, $\phi$ could have a ballistically moving piece; if this existed, one could decouple it and treat it as noise. Focusing on the slow part of $\phi$, the most general equation that one can write has the form
\begin{equation}\label{2mode2}
\partial_t \phi + \partial_x (a\vec{S}\cdot \vec{S} + b\phi^2) = D_\phi \partial_x^2 \phi + \partial_x \zeta + \ldots
\end{equation}
There are two symmetry-allowed terms at this order in fluctuations; other possibilities (e.g., cross terms between $\vec S$ and $\phi$) manifestly break symmetries. We invoke integrability to argue that $\phi$ has a piece that is conserved---if we interpret $\phi$ as related to an energy current, this is intuitive. Finally, we consider the dynamics of a fluctuation of $\phi$ that is unaccompanied by a spatial gradient of $\vec S$. Intuitively, a local imbalance between left- and right-movers ``dissolves'' instead of propagating either left or right, suggesting that $b = 0$. In fact, general properties of GHD forbid any normal mode of the GHZ equations from having a KPZ nonlinearity, constraining $b = 0$~\cite{vir2019}.

The system of equations~(\ref{2mode1}, \ref{2mode2}) with $b = 0$ defines a two-mode hydrodynamic theory for spin fluctuations. One can check that the infinite-temperature state is a stationary measure under these equations~\cite{2022arXiv221203696D}. We have not found an analytic solution for these equations, but numerically we have checked that they reproduce the KPZ scaling function for dynamical correlations, as well as a symmetric distribution for $Q(t)$. However, it is not entirely clear if one can stop at two modes: large-scale simulations of the FCS in classical spin chains~\cite{2003.05957,PhysRevLett.128.090604} suggest a value of the kurtosis that seems to disagree with the two-mode hydrodynamics. Addressing these discrepancies is an important question for future numerical and experimental work.

\subsection{Transport from nonequilibrium conditions}

So far we have considered correlation functions in spatially homogeneous systems. To measure transport, either numerically or experimentally, a standard approach is to create an initial state with a density imbalance and observe the relaxation of that imbalance. This setup has the virtue of only requiring one to measure expectation values, which are simpler to observe than correlations. The most studied example of an imbalanced initial state is the ``domain-wall'' initial condition, for an infinite spin chain where the spins left (right) of the origin are at chemical potential $-\mu (+\mu)$. We expect to recover linear response as $\mu \to 0$, but as we will see the linear-response limit $\mu \to 0$ and the late-time limit $t \to \infty$ do not commute. Indeed, the FCS in the \emph{anisotropic} XXZ model undergoes a phenomenon akin to spontaneous symmetry breaking, where the late-time limit distribution at any $\mu \neq 0$ is universal and $\mu$-independent~\cite{2022arXiv220309526G}.

The initial state can be written as $\rho_\mu \propto \prod\nolimits_i (1 + \mathrm{sign}(i) \mu S^z_i)$. Expanding to first order in $\mu$ and evaluating the Heisenberg-evolved operator $S^z_j(t)$ in this state, we get
\begin{equation}
\langle S^z_j(t) \rangle_\mu \approx \mu \sum_i \mathrm{sign}(i) \mathrm{Tr}(S^z_j(t) S^z_i) \Rightarrow \lim_{\mu \to 0} \frac{\langle S^z_{j-1}(t) \rangle_\mu - \langle S^z_{j}(t)\rangle_\mu}{\mu} = \mathrm{Tr}(S^z_j(t) S^z_0(0)).
\end{equation}
Thus, if the limit $\mu \to 0$ is taken at finite time, this procedure faithfully gives the equilibrium correlation function at that time; to get linear-response transport, one should take the limit $\mu \to 0$ \textbf{before} $t \to \infty$. 

Let us now discuss what happens if we take the late-time limit at $\mu \neq 0$. We first consider the easy-axis regime, in which the analysis is simpler. In the easy-axis regime, one can create the domain-wall initial condition by beginning with an initial state at some fixed global $\mu$, then inserting a giant quasiparticle that occupies half the system. At an $O(1)$ distance from the domain wall, this initial state coincides with $\rho_\mu$. From this construction, we can see that the quasiparticle content of $\rho_\mu$ is as follows: for any finite $s$, the distribution of $s$-strings is spatially homogeneous. Strings with $s \gg 1/\mu$ are exponentially suppressed. When an $s$-string enters or exits the giant quasiparticle, it makes that quasiparticle jitter. As detailed in Ref.~\cite{2022arXiv220309526G} this gives rise to diffusive transport on average, but with strongly enhanced fluctuations. Fluctuations are enhanced, in effect, because the charge carrier has infinite charge, so the shot noise from it is divergent. The FCS in this problem was recently exactly computed for all $\mu$; it takes a universal non-gaussian limit shape for $\mu > 0$, independent of $\mu$, and another (also non-gaussian) limit shape for $\mu = 0$. At small $\mu$, the crossover timescale from $\mu = 0$ behavior to $\mu > 0$ behavior occurs at $t(\mu) \sim 1/\mu^4$. 

Many of these considerations carry over to the Heisenberg point, with one crucial difference: a domain wall between two states of opposite magnetization no longer has $O(1)$ overlap with a single quasiparticle state. Rather, the domain wall has a nontrivial expansion in terms of quasiparticles, as one can see from the fact that the pure domain wall initial state $\mu \to \infty$ actually relaxes (approximately) diffusively~\cite{PhysRevB.99.140301,1742-5468-2017-10-103108}. However, it is easy to see that the domain wall state is at zero energy density, and therefore has zero density of any finite-$s$ strings.
Therefore, to extract the quasiparticle content, it suffices to count bulk quasiparticles; these are suppressed for $s > 1/\mu$ so the superdiffusive dynamics freezes out on the timescale $t(\mu) \sim 1/\mu^3$. Beyond this timescale, we expect the domain wall to relax diffusively, both via jitter from the quasiparticles and via its own intrinsic $z = 2$ ``unwinding'' dynamics. So far, however, we do not have a \emph{hydrodynamic} theory of the relaxation of the $\mu \to \infty$ domain wall: until we have this, it seems that direct theoretical calculations of FCS in this setup remain out of reach.

 In addition to domain walls, recent experiments~\cite{bloch2014,Jepsen:2020aa} have explored the relaxation of spin-spiral initial states. These initial states are product states, in which the orientation of the spin rotates around the XZ plane with a wavelength $\lambda$. More precisely, the initial state is $|\psi\rangle = \prod_i |+\theta_i\rangle$, where $\theta_i = (\sin 2\pi i /\lambda, 0, \cos 2\pi i/\lambda)$ is a vector that rotates in the XZ plane with wavelength $\lambda$, and the state $|+\theta_i\rangle$ describes a spin pointing along the direction $\theta_i$ on the Bloch sphere. 

 Relaxation from spin spirals was recently described within the GHD framework~\cite{Jacopoetal}, \emph{except} at the Heisenberg point. At the Heisenberg point, in the simplest (local density) approximation, GHD predicts that spin spirals do not relax ballistically. This is consistent with the experimental observation that spirals in fact relax approximately ``diffusively,'' i.e., the time for a spiral of wavelength $\lambda$ to relax is $t(\lambda) \sim \lambda^2$. At this point, however, we lack a direct calculation of this relaxation rate (or indeed of the functional form of the relaxation).

\section{EFFECTS OF INTEGRABILITY-BREAKING}

So far, we have focused on the exactly integrable limit. This limit is fine-tuned, and many features (such as exact conserved quantities) are presumably lost the moment one perturbs an infinite system away from this limit. (Even in classical systems, the KAM theorem does not suggest that integrable systems are stable in the thermodynamic limit~\cite{BASKO20111577}.) Given the key part that stable quasiparticles have played in our analysis so far, it would be natural to expect that anomalous transport crosses over to regular transport on some characteristic integrability-breaking timescale. While this is clearly seen for \emph{some} classes of integrability-breaking perturbation, other classes of perturbation have surprisingly robust signatures of anomalous transport. These signatures are seen not only in quantum spin chains (where exact time evolution is limited to very short times), but also in classical spin chains that can be simulated for many thousands of interaction timescales. 

\subsection{Numerics on classical spin chains}

We briefly summarize the numerical results obtained in Ref.~\cite{roy2022robustness} -- see also~\cite{2022PhRvB.105j0403M}. That work considers the dynamics of a chain of classical spins (i.e., degrees of freedom living on the 2-sphere) governed by the Hamiltonian 
\begin{equation}
H = - \sum_i \left(\log(1 + \vec{S}_i \cdot \vec{S}_{i+1}) + \lambda \vec{S}_i \cdot \vec{S}_{i+1} \right).
\end{equation}
When $\lambda = 0$, this model is integrable; $\lambda$ is the scale of the integrability-breaking perturbation. When $\lambda = 0.5$, Ref.~\cite{roy2022robustness} finds that energy transport is clearly diffusive, while spin transport remains superdiffusive with autocorrelation functions consistent with KPZ. These observations persist for times that exceed the natural precessional timescales by a factor of nearly a thousand. These observations suggest, at the very least, a decoupling between the lifetimes of quasiparticles responsible for spin and energy transport. Related work on nonintegrable classical spin chains has shown the persistence of large soliton-like excitations away from integrability~\cite{PhysRevE.106.L062202}, though at present no concrete proposal relates these excitations to high-temperature superdiffusion. 

\begin{figure}
 \label{fig:IntBreaking} 
\begin{centering}	\includegraphics[width=0.8\textwidth]{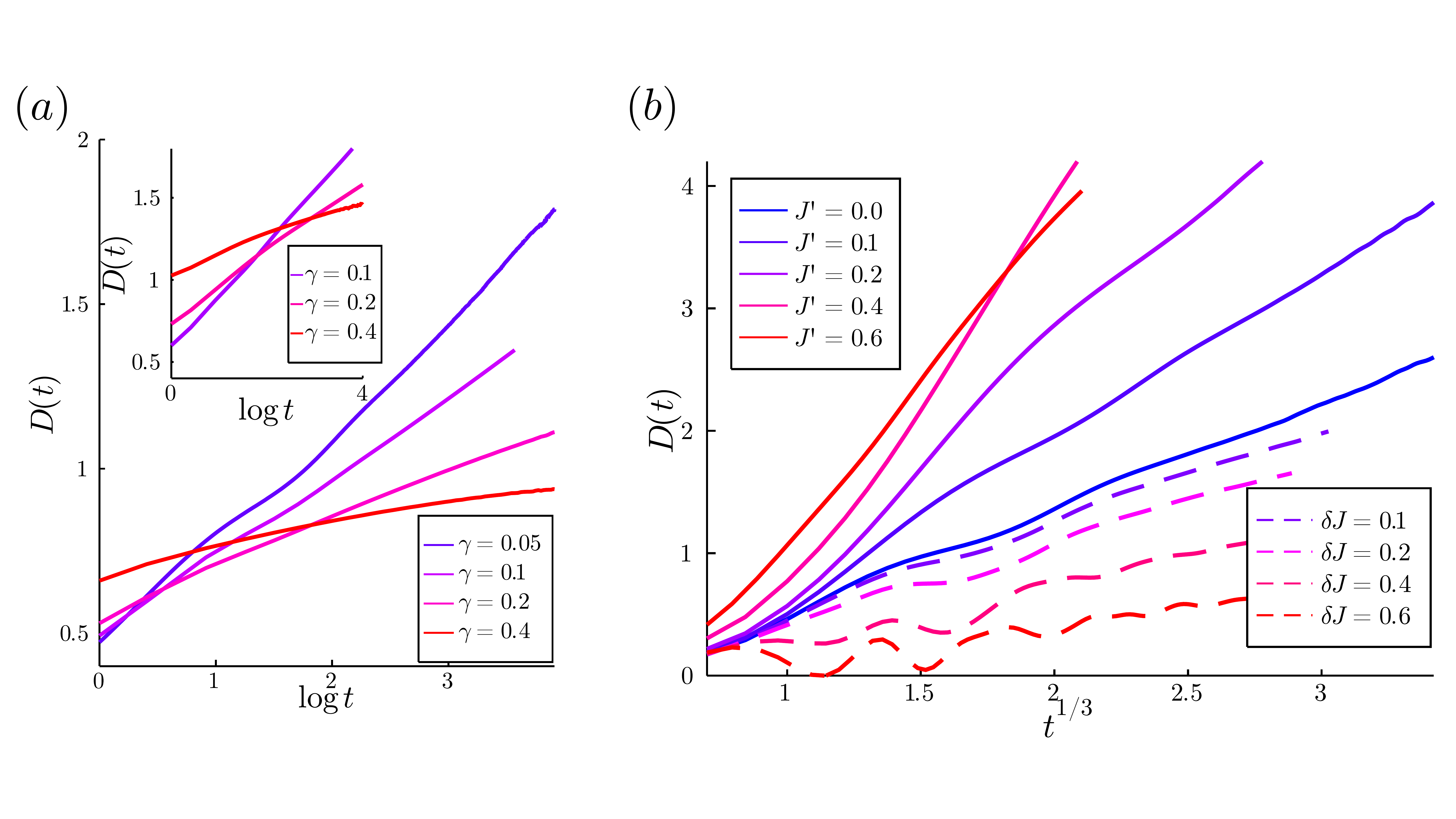}
\end{centering}
	\caption{ {\bf Breaking integrability and stability of superdiffusion} -- (a) Time-dependent diffusion constant for Heisenberg magnets with $SU(2)$-symmetric classical noise (main figure: noise coupling to energy, inset: noise coupling to energy current), showing logarithmic diffusion for weak noise strength $\gamma$. (b) Time-dependent diffusion constant for Heisenberg magnets with $SU(2)$-symmetric energy-preserving perturbations (staggered energy couplings and next-nearest-neighbor interactions). Transport remains superdiffusive with exponent $z=3/2$ for all accessible time scales. 
 Figure reproduced from Ref.~\cite{PhysRevLett.122.210602}.
            }             
\end{figure}

\subsection{Golden Rule estimates}

Spin transport is due to the largest quasiparticles and energy transport to the smallest. Therefore, if for some reason larger quasiparticles persist for longer, this would explain (at least qualitatively) the persistence of anomalous spin transport in a regime with energy diffusion. In fact there is an obvious mechanism for larger quasiparticles to live longer: when the integrability-breaking perturbation conserves $SU(2)$ symmetry, it cannot efficiently scatter long-wavelength Goldstone excitations associated with that symmetry: locally, these excitations look like rotated vacua, and as such the perturbation does not directly couple to them. Instead, $SU(2)$-respecting perturbations couple to $s$-strings with a matrix element that is suppressed as $1/s$, and therefore a rate that is suppressed at least as $1/s^2$~\cite{PhysRevLett.127.057201}. In the case of $SU(2)$-conserving noise, this perturbative analysis can be controlled and we find $\tau_s \sim s^2$, giving an a.c. conductivity that diverges only logarithmically at low frequencies. For $SU(2)$-conserving Hamiltonian perturbations, the kinematics is much more complicated, and we do not have a reliable estimate of $\tau_s$. At the level of low-order perturbation theory it is plausible that $\tau_s > s^3$, in which case superdiffusion would persist despite the lack of strictly conserved quantities. Numerical results are in good agreement with these perturbative predictions (Fig.~\ref{fig:IntBreaking}).

Could superdiffusion really be the asymptotic behavior? There are three possibilities as a function of the integrability-breaking parameter $\lambda$: superdiffusion persists for all $\lambda$, superdiffusion is asymptotically absent for any $\lambda \neq 0$, or there is a phase transition between superdiffusion and regular diffusion at some critical value of $\lambda$. The third scenario cannot be ruled out but seems highly implausible, as it would require a new mechanism to ``turn on'' at some finite $\lambda$. The first scenario \emph{can} be ruled out, as the limit of large noise and weak Hamiltonian couplings is analytically tractable~\cite{PhysRevLett.128.246603, 2020arXiv200713753G}. This leaves us with the second scenario---superdiffusion is asymptotically lost~\cite{2020arXiv200713753G}, but the crossover to diffusion might require processes at very high orders in perturbation theory, so it might be very high-order or even nonperturbative at small $\lambda$. However, beyond simple estimates, a full perturbative theory of the effects of integrability-breaking perturbations remains to be developed~\cite{friedman2019diffusive,bastianello2020generalised,2020arXiv200411030D,Bastianello_2021}.

\subsection{``Weak'' integrability breaking}

Finally, we note the recently developed idea of ``weak integrability breaking''~\cite{PhysRevB.105.104302, szasz2021weak, 2023arXiv230212804S, orlov2023adiabatic}. Consider an integrable Hamiltonian $H$. By a local unitary transformation, $H$ can be transformed into the isospectral integrable Hamiltonian $H' = U^\dagger H U$. Expanding $U \approx 1 + \epsilon V$, $H' \approx H + \epsilon [H,V]$. The perturbation $[H,V]$ is called a ``weak perturbation,'' since (at leading order) it yields an integrable model, and therefore does not cause local observables to relax. Surprisingly, many commonly considered integrability-breaking perturbations, such as next-nearest-neighbor couplings in the Heisenberg model, belong to this class of weak perturbations. Currently our understanding of such weak perturbations remains rudimentary. 
It is unclear if they can be relevant to the observations in integrable classical spin chains: weak perturbations are just as incapable of relaxing energy as of relaxing spin, so it does not seem that they can explain the distinction between energy and spin relaxation timescales.


\section{SUMMARY AND OUTLOOK}

In this article we have attempted to summarize the current understanding of anomalous transport in the Heisenberg spin chain---and more generally in any integrable one-dimensional lattice model invariant under a continuous nonabelian symmetry. All such models have large-scale classical solitons that one can regard as slow rotations of the orientation of the ferromagnetic vacuum. Starting from the ferromagnetic vacuum, one can construct a thermal state by assembling solitons of all sizes. (This is true regardless of whether the couplings are ferromagnetic or antiferromagnetic.) Since the model is integrable, these solitons remain stable at finite density. The interactions between solitons at different scales give rise to the singular screening effects that (as we discussed) underlie anomalous transport. 

At the linear response level, we have analytic calculations of the transport exponent $z = 3/2$, as well as compelling numerical evidence that the dynamical spin correlation functions (and more generally, the autocorrelators of densities that transform nontrivially under the nonabelian symmetry) have the same functional form as the density correlation function in the Burgers equation. Regardless of this correspondence, the spin density cannot obey a Burgers equation as the latter is inherently chiral. A recent proposal attempts to reconcile these observations by invoking a two-component fluctuating hydrodynamics that respects all the symmetries of the spin chains and reproduces the dynamical two-point functions. The predictions of this two-component theory beyond linear response remain to be tested. Fortunately, present-day experiments in ultracold atomic gases and superconducting qubit arrays allow access to quantities beyond linear response---such as full counting statistics---that allow for much more sensitive tests of these novel transport scenarios. 

As some puzzles have moved toward resolution, others have taken their place. Of these, perhaps the most intriguing is the robustness of superdiffusive spin transport away from integrability, at least for integrability-breaking perturbations that respect the underlying nonabelian symmetry. The experimental evidence for this behavior is compelling, but we lack an adequate theoretical framework for discussing integrability-breaking. The surprising recent discovery of weak integrability-breaking perturbations suggests that the progress that remains to be made on this front includes not just better quantitative methods, but perhaps also new concepts.

\section*{DISCLOSURE STATEMENT}
 The authors are not aware of any affiliations, memberships, funding, or financial holdings that might be perceived as affecting the objectivity of this review. 

\section*{ACKNOWLEDGMENTS}

We thank Utkarsh Agrawal, Vir Bulchandani, Jacopo De Nardis, Benjamin Doyon, Fabian Essler, Michele Fava, David Huse, Enej Ilievski, Vedika Khemani, Alan Morningstar, Vadim Oganesyan, Sid Parameswaran, Toma\v{z} Prosen, Marcos Rigol, Subir Sachdev, Brayden Ware, David Weiss, and Marko \v{Z}nidari\v{c} for collaborations and/or discussions on topics related to this review.
This work was supported by the National Science Foundation under NSF Grant No. DMR-1653271 (S.G.), the US Department of Energy, Office of Science, Basic Energy Sciences, under Early Career Award No. DE-SC0019168 (R.V.), and the Alfred P. Sloan Foundation through a Sloan Research Fellowship (R.V.).

%
\bibliographystyle{unsrt11}

\bibliography{refs}

\end{document}